\documentclass[prl,nofootinbib,superscriptaddress,twocolumn]{revtex4}
\usepackage[a4paper,left=1.5cm,right=1.5cm,top=3cm,bottom=3cm]{geometry}

\usepackage{amssymb,amsmath,amsfonts}
\usepackage[dvipsnames]{xcolor}
\usepackage{graphicx}
\usepackage{longtable}
\usepackage{verbatim}
\usepackage{color}
\usepackage{mdframed}
\usepackage{soul}
\usepackage{amsfonts,amssymb,mathrsfs,amsmath,esint}
\usepackage{slashed, cancel}
\usepackage{framed}
\usepackage{mdframed}
\usepackage{simplewick} % For Wick contractions
\allowdisplaybreaks

\usepackage{latexsym}
\usepackage{graphicx}
\graphicspath{{./Figures/}}
\usepackage[dvipsnames]{xcolor}
\usepackage{booktabs}
\usepackage{datetime}
\newdateformat{mydate}{\THEDAY{ }\monthname[\THEMONTH]{ }\THEYEAR}

\usepackage{tikz}
\usepackage{color}
\usepackage{framed}
\usepackage{hyperref}
\hypersetup{colorlinks, citecolor=bluscuro, linkcolor=black, urlcolor=bluscuro}
\definecolor{rossos}{cmyk}{0,1,1,0.55}
\definecolor{bluscuro}{rgb}{0.15, 0.2, .85}
\definecolor{bluchiaro}{cmyk}{1,.3,0.,0.1}

\definecolor{mygreen}{RGB}{0,130,0}

\graphicspath{{./Figures/}}

\newcommand{\be}{\begin{equation}}
\newcommand{\ee}{\end{equation}}
\newcommand{\bea}{\begin{eqnarray}}
\newcommand{\eea}{\end{eqnarray}}
\newcommand{\beq}{\begin{equation}}
\newcommand{\eeq}{\end{equation}}
\def\beqa{\begin{eqnarray}}

\def\eeqa{\end{eqnarray}}

\def\lsim{\mathrel{\rlap{\lower4pt\hbox{\hskip0.5pt$\sim$}}
    \raise1pt\hbox{$<$}}}         %less than or approx. symbol
\def\gsim{\mathrel{\rlap{\lower4pt\hbox{\hskip0.5pt$\sim$}}
    \raise1pt\hbox{$>$}}}         %greater than or approx. symbol

\newcommand{\ud}{\mathrm{d}}

\usepackage[normalem]{ulem}
\usepackage{soul}

\begin{document}

\title{Circular Polarization of the Astrophysical Gravitational Wave Background}

\author{L. Valbusa Dall'Armi}
\address{Dipartimento di Fisica e Astronomia ``G. Galilei",
Universit\`a degli Studi di Padova, via Marzolo 8, I-35131 Padova, Italy}
\address{INFN, Sezione di Padova,
via Marzolo 8, I-35131 Padova, Italy}

\author{A. Nishizawa}
\affiliation{Research Center for the Early Universe (RESCEU), School of Science, The University of Tokyo, Tokyo 113-0033, Japan}

\author{A. Ricciardone}

\address{Dipartimento di Fisica ``E. Fermi'', Universit\`a  di Pisa, I-56127 Pisa, Italy}

\address{Dipartimento di Fisica e Astronomia ``G. Galilei",
Universit\`a degli Studi di Padova, via Marzolo 8, I-35131 Padova, Italy}

\address{INFN, Sezione di Padova,
via Marzolo 8, I-35131 Padova, Italy}

\author{S. Matarrese}
\address{Dipartimento di Fisica e Astronomia ``G. Galilei",
 Universit\`a degli Studi di Padova, via Marzolo 8, I-35131 Padova, Italy}

\address{INFN, Sezione di Padova,
via Marzolo 8, I-35131 Padova, Italy}

\address{INAF - Osservatorio Astronomico di Padova, Vicolo dell'Osservatorio 5, I-35122 Padova, Italy}

\address{Gran Sasso Science Institute, Viale F. Crispi 7, I-67100 L'Aquila, Italy}

\date{\today}

\begin{abstract}
\noindent
The circular polarization of gravitational waves is a powerful observable to test parity violation in gravity and to distinguish between the primordial or the astrophysical origin of the stochastic background. This property comes from the expected unpolarized nature of the homogeneous and isotropic astrophysical background, contrary to some specific cosmological sources that can produce a polarized background. However, in this work we show that there is a non-negligible amount of circular polarization also in the astrophysical background, generated by Poisson fluctuations in the number of unresolved sources, which is present in the third-generation interferometers with signal-to-noise ratio larger than two. We also explain in which cases the gravitational wave maps can be cleaned from this extra source of noise, exploiting the frequency and the angular dependence, in order to search for signals from the early Universe. Future studies about the detection of polarized cosmological backgrounds with ground- and space-based interferometers should account for the presence of such a foreground contribution.
\end{abstract}

\maketitle

\paragraph{Introduction} 
The astrophysical gravitational wave background (AGWB) generated by the superposition of many unresolved signals emitted by stellar-mass binary black holes (BBH)~\cite{Phinney:2001di,Regimbau:2011rp,Zhu:2011bd,Zhu:2012xw} is one of the main target of the third-generation (3G) interferometers Einstein Telescope (ET)~\cite{Maggiore:2019uih} and Cosmic Explorer (CE)~\cite{Evans:2021gyd}. Such AGWB is assumed to be unpolarized in General Relativity, since the $V$ Stokes parameter of the gravitational wave emitted by each binary is washed out by the average over the inclination angle of the system w.r.t. the direction of observation, if such an angle is distributed isotropically. However, the AGWB is not perfectly homogeneous and isotropic, but it exhibits anisotropies of about 1 part in 100, mainly due to the shot noise fluctuation in the number of GW sources~\cite{Jenkins:2019uzp,Jenkins:2019nks,Alonso:2020mva,Bellomo:2021mer}. In this \textit{letter} we compute the amount of circular polarization of the AGWB induced by these Poisson fluctuations, which are expected to be large in light of the low number of BBH detected by LVK in the local Universe~\cite{LIGOScientific:2021psn}. The computation of the circular polarization of the AGWB is similar to the CMB one, because also in the case of photons the only non-vanishing Stokes parameter at the isotropic level is the intensity/temperature, while the polarization of the radiation can be detected by looking at the anisotropies generated by scalar perturbations (E modes) or by weak lensing and primordial GWs (B modes)~\cite{Zaldarriaga:1996xe,Lewis:2006fu}. 

Our main result is that the sky map of the circular polarization of the AGWB has zero average, but non-vanishing fluctuations, which are generated by a flat (in the multipole space) angular power spectrum and which decrease with the observing time as $1/\sqrt{T_{\rm obs}}$. The amount of circular polarization produced by deviations from an isotropic distribution of the sources is present in the network ET+CE with signal-to-noise ratio (SNR) larger than two for any observing time. To observe such a background it is necessary to combine the techniques used to measure an isotropic and circularly polarized stochastic background~\cite{Crowder:2012ik,Domcke:2019zls,Orlando:2020oko,Martinovic:2021hzy} and to observe the anisotropies of stochastic backgrounds~\cite{Allen:1996gp,Cornish:2001hg,Mentasti:2020yyd,Alonso:2020rar,LISACosmologyWorkingGroup:2022kbp}.

\begin{figure}[t!]
\centering 
\includegraphics[width=0.45\textwidth]{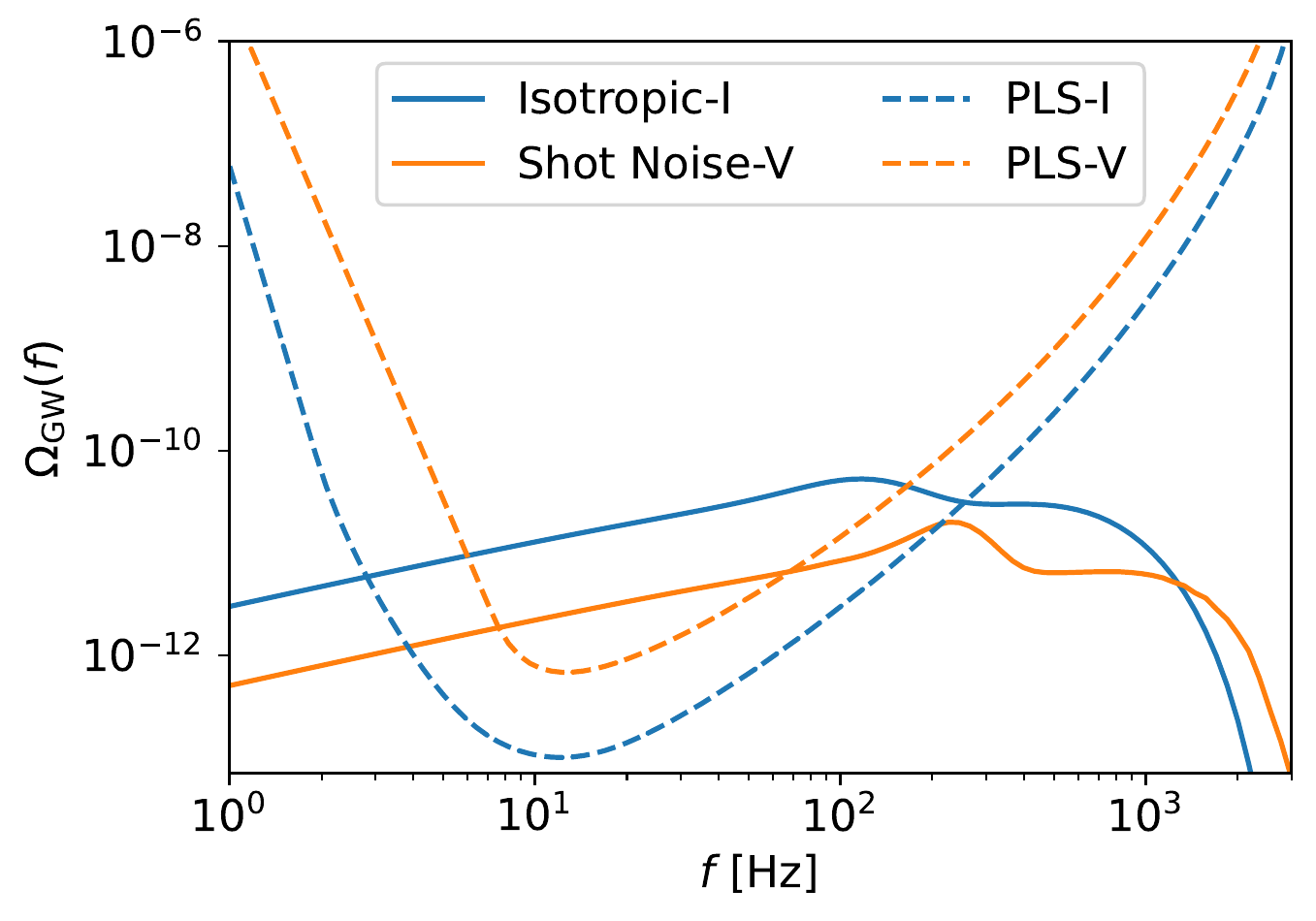}
\caption{\it Plot of the dominant contribution to the intensity (solid blue) and to the circular polarization (solid orange) of the AGWB as a function of the frequency. The contribution to the intensity is isotropic, while the circular polarization is anisotropic, because it is generated by the shot noise of the sources. In the same figure the Power-Law Sensitivity (PLS) curves of the detector network made by one Einstein Telescope and two Cosmic Explorer ($\rm SNR=1$) for the intensity (dashed blue) and circular polarization (dashed orange) for one year of observation are depicted.} 
\label{fig:all_sgwb_terms_and_sensitivities}
\end{figure}

Furthermore, in our work we have shown how the circular polarization of the AGWB constitutes a troublesome foreground for the detection of the V Stokes parameter of stochastic backgrounds of cosmological origin (CGWB), which could be generated by several mechanisms in the early Universe, (e.g., axion inflation~\cite{Anber:2006xt,Barnaby:2010vf,Sorbo:2011rz,Barnaby:2011vw,Barnaby:2011qe}, Chern-Simons operators~\cite{Bartolo:2017szm,Bartolo:2018elp,Bartolo:2020gsh}). More specifically, we have found that the different angular and frequency dependence of the circular polarization of the AGWB and the CGWB allows us to discriminate between the two contributions, although the astrophysical foreground gives non-negligible corrections to the estimate of the cosmological background that need to be properly taken into account.

\paragraph{Characterization of the astrophysical background}

A GW of frequency $f$ is described by the strain $h_{ij}(f,\hat{n}) =  h_{+}(f)e_{ij,+}(\hat{n}) + h_{\times}(f)e_{ij,\times}(\hat{n})$, where $\hat{n}$ is the direction of observation in the sky. Given a direction of observation $\hat{n}$ and a coordinate system $\{\hat{x},\hat{y},\hat{z}\}$ arbitrarily oriented, we define\footnote{When $\hat{n}\parallel \hat{z}$ the unit vectors $\hat{u}$ and $\hat{v}$ stay in the $(\hat{x},\hat{y})$ plane.} the orthonormal basis $\left\{\hat{n},\hat{u},\hat{v}\right\}$ by using $\hat{u}\equiv \hat{n}\times \hat{z}/|\hat{n}\times \hat{z}|$ and $\hat{v}\equiv \hat{n}\times \hat{u}$. These allow us to define the following real polarization tensors~\cite{Bartolo:2018qqn,Domcke:2019zls},
\begin{equation}
    e_{ij,+} \equiv \frac{u_iu_j-v_iv_j}{\sqrt{2}}\;, \hspace{3em}e_{ij,\times} \equiv \frac{u_iv_j+u_jv_i}{\sqrt{2}}\, ,
\end{equation}
which are related to the chiral ones by
\begin{equation} 
        h_R = \frac{h_+-ih_\times}{\sqrt{2}}\; , \hspace{3em} 
        h_L = \frac{h_++ih_\times}{\sqrt{2}}\; .
\end{equation}
The intensity and the circular polarization associated to this wave are defined by using
\begin{equation}
    S_I \equiv |h_{R}|^2+|h_{L}|^2\; , \hspace{2.5em} S_V \equiv  |h_{R}|^2-|h_{L}|^2\; .
    \label{eq:def_Stokes_parameters_single_source}
\end{equation}
We characterize the AGWB in terms of energy density per logarithmic frequency, normalized to the critical energy density of the Universe today, for the intensity and polarization
\begin{equation}
    \Omega_{\rm AGWB}^I \equiv  \frac{4\pi^2 f^3}{3H_0^2}S^{{\rm AGWB}}_I\, , \hspace{1.75em} \Omega_{\rm AGWB}^V \equiv  \frac{4\pi^2 f^3}{3H_0^2}S^{{\rm AGWB}}_V\, ,
	\label{eq:stokes_to_omega}
\end{equation}
where $H_0$ is the value of the Hubble parameter at present and the Stokes parameters have been computed by considering the total AGWB strain, $h_{\rm AGWB}$, which is the sum of the strains emitted by all the unresolved sources. To compute the GW energy density associated to the intensity and polarization we use the prescription given by~\cite{Phinney:2001di,Regimbau:2011rp},
\begin{equation}
	\begin{split}
	\Omega_{\rm AGWB}^\alpha(\hat{n},f) = \frac{f}{\rho_{\rm crit} c^2} \sum_{t_d,		\vec{\theta},z} &\, \frac{N_{\rm GW}(\hat{n},t_{d},\vec{\theta},z)\frac{\ud 		E^\alpha}{\ud f_e\ud\Omega_e}\left(\vec{\theta},z,f\right)}{(1+z)H(z)T_{\rm obs}		\frac{\ud V}{\ud z\ud 	\Omega_e}(z)}
	\label{eq:omega_stokes_agwb_basic}
	\end{split}
\end{equation}
where $\alpha$ identifies the Stokes parameters $I$ and $V$, $\vec{\theta}$ represents the intrinsic parameters of a binary, $\Omega_e$ is the solid angle at the emitter, $z$ is the redshift at which the GWs are emitted, and $t_d$ the time-delay between the formation of the binary and the merger~\cite{Fishbach:2021mhp}. We have discretized the parameter space in small bins of width $\Delta t_d$, $\Delta \vec{\theta}$ and $\Delta z$. $N_{\rm GW}(\hat{n},t_{d},\vec{\theta},z)$ is the number of unresolved BBH mergers in the observation time $T_{\rm obs}$, characterized by the parameters $\vec{\theta}$ in the comoving volume along the direction $\hat{n}$ at redshift $z$ and formed at $t(z)-t_d$. Additional details about the population of BBHs can be found in the Supplemental Material, which includes also Refs.~\cite{LIGOScientific:2020kqk,Behroozi:2019kql,Bellomo:2021mer,Tinker:2008ff,Mapelli:2017hqk,Mapelli:2019bnp}. We have also introduced the ``polarized energy spectrum''
\begin{equation}
\begin{split}
    \frac{dE^\alpha}{df_e d\Omega_e}(\vec{\theta},f_e)\Bigl|_{f_e = (1+z)f} \equiv& \frac{\pi d_L^2(z) c^3 f^2}{2G(1+z)^2} S_\alpha(\vec{\theta},f)\\
    =& A_E\left(m_1,m_2,\chi_1,\chi_2,z,f\right) Q_\alpha(\iota) \, ,
    \label{eq:dE_df_Stokes}
\end{split}
\end{equation}
with $\chi_1$, $\chi_2$ the spin of the BHs in the binary, $d_L$ the luminosity distance of the source, $G$ the gravitational constant and $S_\alpha$ defined in Eq. \eqref{eq:def_Stokes_parameters_single_source}. The dependence on the inclination angle of the binary $\iota$ is encoded in the factor~\cite{Peters:1963ux,Finn:1992xs,Cutler:1994ys,Maggiore:2007ulw}
\begin{equation}
    Q_\alpha(\iota)\equiv \begin{cases}
    \left(\frac{1+\cos^2\iota }{2}\right)^2+\cos^2\iota \hspace{3.5em} \alpha = I \\
    ( 1+\cos^2\iota )\cos\iota \hspace{5em} \alpha = V
    \end{cases}
    \label{eq:def_Q_alpha}
\end{equation}

 We consider all the events with $\rm SNR \leq 20$ for each of the interferometers of the network as unresolved sources. This very conservative choice has been made to avoid errors in the estimate of the AGWB amplitude due to the subtraction of the resolved sources~\cite{Sachdev:2020bkk,Zhou:2022otw,Zhou:2022nmt,Zhong:2022ylh}. The important thing is that the number of GW events depends on the inclination angle through 
\begin{equation}
	N_{\rm GW}(\hat{n},t_{d},\vec{\theta},z) \sim p(\iota) = \frac{\sin\iota}{2}\, ,
\end{equation}
where we have considered a uniform distribution for the inclination angle of the binary. The homogeneous and isotropic contribution to the AGWB depends on the following integrals over the inclination angle of the binary,
\begin{eqnarray}
\bar{\Omega}_{\rm AGWB}^I &\sim& \int d\iota \, p(\iota )Q_I(\iota) = \frac{4}{5} \, ,\nonumber  \\
\bar{\Omega}_{\rm AGWB}^V &\sim& \int d\iota \, p(\iota )Q_V(\iota)  = 0 \, ,
\label{eq:V_iso_iota_contribution}
\end{eqnarray}
with $Q_\alpha(\iota)$ defined in Eq. \eqref{eq:def_Q_alpha}. This means that the isotropic distribution of sources washes out the circular polarization of the single sources, which sum incoherently, as expected~\cite{Regimbau:2011rp}.

\paragraph{Shot-noise contribution to the circular polarization}
However, the astrophysical background is produced by the superposition of the finite number of unresolved signals, therefore it is affected by shot-noise. More specifically, the number of GW events $N_{\rm GW}$ that contribute to the background is obtained as the sum of the GW events over all halos in the Universe. Both the number of halos and the GW events in a single halo fluctuate following a Poisson distribution, therefore the total number of mergers is described by a Compound Poisson Distribution (CPD)~\cite{Jenkins:2019uzp,Jenkins:2019nks}. The expectation value of the CPD, $\bar{N}_{\rm GW}$, is used in Eq. \eqref{eq:omega_stokes_agwb_basic} to compute the homogeneous and isotropic contribution to the background, $\bar{\Omega}_{\rm AGWB}^\alpha$, which results in the integrals of Eq. \eqref{eq:V_iso_iota_contribution} over the inclination angle. In the Supplemental Material, we show that the covariance associated to the shot-noise fluctuation is 
\begin{equation}
    \begin{split}
        C_{f f^\prime,\hat{n}\hat{n}^\prime}^{\alpha \alpha^\prime}\equiv&\,{\rm cov}\left[\Omega_{\rm AGWB}^\alpha(\hat{n},f),\Omega_{\rm AGWB}^{\alpha^\prime}(\hat{n}^\prime,f^\prime)\right]  \\
        =&\sum_{t_d,\vec{\theta},z}\frac{1}{\left[(1+z)H(z)\right]^2}\frac{ff^\prime}{(\rho_{\rm crit}c^2)^2}\bar{N}_{h}(M_{h}, t_{d}, z)\\
        &\times\frac{\frac{\ud E^\alpha}{\ud f_e \ud\Omega_e}\left(\vec{\theta},z,f\right)\frac{\ud E^{\alpha^\prime}}{\ud f_e \ud\Omega_e}\left(\vec{\theta},z,f^\prime\right)}{\left[T_{\rm obs}\frac{\ud V}{\ud z\ud \Omega_e}(z)\right]^2}\frac{\delta_{\hat{n}\hat{n}^\prime}}{\Delta\Omega}\\
        &\times\biggl[\bar{N}_{\rm GW|h}(M_{h},t_{d},\vec{\theta},z)+\bar{N}^2_{\rm GW|h}(M_{h},t_{d},\vec{\theta},z)\biggl] \, ,
    \end{split}
    \label{eq:cv_pol_sn}
\end{equation}
where $\bar{N}_h$ is the average number of halos with mass $M_h$ in the Universe at time $t(z)-t_d$, while $\bar{N}_{\rm GW|h}$ is the average number of GW sources per halo of mass $M_h$ with parameters $\vec{\theta}$ produced at $t(z)-t_d$. $\Delta\Omega$ is the solid angle of the pixels in which we have divided the sky, $\Delta\Omega \equiv 4\pi/N_{\rm pix}$. The two contributions to the shot-noise match the results of~\cite{Jenkins:2019uzp,Jenkins:2019nks,Alonso:2020mva} and, in agreement with~\cite{Bellomo:2021mer}, we have found that the $N^2_{\rm GW|h}$ term is subdominant. The covariance of the shot-noise is proportional to $\delta_{\hat{n}\hat{n}^\prime}$, which means that the angular power spectrum of the signal is the same at all  multipoles~\cite{Jenkins:2019uzp,Jenkins:2019nks,Alonso:2020mva}. In this case, the sum over all possible inclination angles of the binaries does not give a vanishing contribution to the circular polarization of the astrophysical background as in Eq. \eqref{eq:V_iso_iota_contribution}, because the integral over the inclination angle that determines the amplitude of the fluctuations is quadratic in $Q_\alpha$ defined in Eq. \eqref{eq:def_Q_alpha},
\begin{equation}
\begin{split}
C_{f f^\prime,\hat{n}\hat{n}^\prime}^{\alpha \alpha^\prime} \sim& \int d\iota \, p(\iota )Q_\alpha(\iota)Q_{\alpha^\prime}(\iota) \rightarrow \begin{cases}
C_{f f^\prime,\hat{n}\hat{n}^\prime}^{I I} \sim \frac{284}{315} \\
C_{f f^\prime,\hat{n}\hat{n}^\prime}^{I V}=0\\
C_{f f^\prime,\hat{n}\hat{n}^\prime}^{V V}\sim \frac{92}{105}
\label{eq:sn_iota_contributions}
\end{cases}
\end{split}
\end{equation}
Since the intensity of the AGWB is dominated by the homogeneous and isotropic contribution, in this work we do not focus on the shot noise fluctuation in the intensity, although an amplitude close to the one of the circular polarization is expected, because $(92/105)/(284/315) \approx 0.972$.

Equation \eqref{eq:cv_pol_sn} shows that the correlation of the shot-noise at different frequencies, for $f\lesssim 100\, \rm Hz$, is equal to one, because $\Omega_{\rm AGWB}^\alpha(\hat{n},f) \sim f^{2/3}$, thus any dependence on the frequency can be factored out. This means that the circular polarization of the AGWB at $f_{\rm piv}$ univocally determines the amplitude of the fluctuations at $f\lesssim 100\, \rm Hz$, because the signal scales as $(f/f_{\rm piv})^{2/3}$. As we will see in the next paragraph, the spectral tilt predicted by the binary evolution is a crucial ingredient to detect (and to subtract) the intensity and the circular polarization of the AGWB. 

There is a large amount of uncertainty in the covariance of the circular polarization due to shot-noise, for example, coming from the BBH merger rate and from the distribution of the intrinsic parameters of the binaries. The shot-noise fluctuation is also sensitive to the number of sources that the detector network can resolve; thus the shot-noise changes if the sensitivity of the instruments changes and if a different SNR threshold of the single source is chosen. The shot-noise scales indeed with $\sqrt{N_{\rm GW}}$, while the homogeneous and isotropic term depends linearly on $N_{\rm GW}$. Note also that in this work we have not focused on the intrinsic~\cite{Namikawa:2015prh,Namikawa:2016edr,Cusin:2017fwz,Cusin:2018rsq,Jenkins:2018kxc,Bertacca:2019fnt} and kinetic~\cite{Cusin:2022cbb,ValbusaDallArmi:2022htu,Chung:2022xhv} anisotropies, induced by GR corrections and by our peculiar motion w.r.t. the sources respectively, because these effects are expected to be negligible if compared to the shot noise~\cite{Bellomo:2021mer,ValbusaDallArmi:2022htu}.

\paragraph{Detection with 3$G$ interferometers}

The detectability of the anisotropies of the intensity of the AGWB, including the shot-noise, has been already discussed in~\cite{Bellomo:2021mer}, therefore in this section we focus on the detectability of the circular polarization of the AGWB, marginalizing w.r.t. the intensity (monopole + shot-noise) of the AGWB. In analogy with the computations of~\cite{Domcke:2019zls,Orlando:2020oko,Martinovic:2021hzy}, it can be shown that the covariance matrix associated to the determination of the shot-noise map, $\Omega_{\rm AGWB}^V(\hat{n},f)$, is
\begin{equation}
  \mathcal{COV}^{VV}_{\hat{n}\hat{n}^\prime} = \frac{1}{T_{\rm obs}}\frac{2\delta_{\hat{n}\hat{n}^\prime}}{\sum_f {\rm Tr}\left(B^V_{f,\hat{n}}S_f^{-1}B^V_{f,\hat{n}}S_f^{-1}\right)}\, ,
  \label{eq:cov_sn_est}
\end{equation}
where
\begin{equation}
    B_{f,\hat{n}}^{{\rm V},AB} \equiv \Delta \Omega \frac{3H_0^2}{4\pi^2 f^3}  \left( F_{A,f\hat{n}}^+ F_{B,f\hat{n}}^{\times\,*}-F_{A,f\hat{n}}^\times F_{B,f\hat{n}}^{+\,*}\right)\,i \frac{f^{2/3}}{f^{2/3}_{\rm piv}}\, ,
    \label{eq:def_B_V}
\end{equation}
with $f_{\rm piv}=1\, \rm Hz$ in our case and $F_A$ the pattern function of the detector $A$ along the direction of observation $\hat{n}$. The matrix $S_f$ is proportional to the covariance matrix of the data,
\begin{equation}
S_f^{AB} \equiv N^{AB}_f+\sum_{\hat{n}^\prime}\sum_\alpha B^{\alpha,AB}_{f, \hat{n}^\prime}\Omega_{\rm piv,\hat{n}^\prime}^{\alpha,\rm SN}\, ,
\label{eq:def_S_f}
\end{equation}
with $N^{AB}_f$ the power spectrum of the noise of the interferometers and $\Omega_{\rm piv,\hat{n}^\prime}^{\alpha,\rm SN}$ the realization of the shot-noise we would like to measure at $f=f_{\rm piv}$. The value of the SNR depends on the specific realization of the shot-noise of the circular polarization\footnote{The angular power spectrum of the shot noise is constant in the multipole space, thus all the first few multipoles give non-negligible contributions to the SNR.}, but we know that on average it is equal to 
\vskip .3 truecm
\begin{equation}
    \langle {\rm SNR} \rangle = \sqrt{\sum_{\hat{n},\hat{n}^\prime}C_{f_{\rm piv} f_{\rm piv},\hat{n}\hat{n}^\prime}^{VV}\left(\mathcal{COV}^{-1}\right)^{VV}_{\hat{n}\hat{n}^\prime}}\, .  
\end{equation}
\vskip .3 truecm
We have computed the SNR for the detector network assuming ET in a 10-km arm triangular configuration placed in Sardinia~\cite{ETsens}, a L-shape 40km CE in Hanford, and 20km CE in Livingston~\cite{CEsens}. We find $\langle \rm SNR \rangle \approx 2.2$ for any observing time, %\an{Should this be twice due to the correction of a factor of four for the signal power spectrum?} 
showing that the circular polarization of the AGWB can be a target for the 3G detectors. In Figure \ref{fig:all_sgwb_terms_and_sensitivities} we have plotted the result of the ‘‘mean amplitude'' of the fluctuation in the circular polarization and the intensity of the AGWB and we have compared them with the Power-Law Sensitivity (PLS) curves for the network ET+CE. The PLS for the polarized background extends the result obtained in~\cite{Thrane:2013oya} to the case of an anisotropic background with constant angular power spectrum. Furthermore, a joint estimate of the AGWB anisotropies with other Stokes parameters (as the intensity), would increase the SNR and could be used to constrain the BBH population too, as we will discuss in the next paragraph.
\paragraph{Foreground subtraction}
We have shown that the circular polarization of the AGWB constitutes a target for the 3G interferometers, but, at the same time, it could also be a limitation for the detection of polarized cosmological signals. For simplicity, we consider a homogeneous and isotropic cosmological background characterized by a power-law spectrum of the following form,
\begin{eqnarray}
     \bar{\Omega}_{\rm CGWB}^I(f) &=& \bar{A}^I_{\rm CGWB} \, \left(\frac{f}{1\, \rm Hz}\right)^{n_T^I}\, , \nonumber\\
       \bar{\Omega}_{\rm CGWB}^V(f) &=& \bar{A}^V_{\rm CGWB} \,\left(\frac{f}{1\, \rm Hz}\right)^{n_T^V}\, ,
\end{eqnarray}
where $\bar{A}^I_{\rm CGWB}$, $\bar{A}^V_{\rm CGWB}$ (the amplitudes of the background at $1\, \rm Hz$), $n_T^I$ and $n_T^V$ are the cosmological parameters we would like to estimate. In this \textit{letter}, we show an example of foreground removal for $n_T^I=n_T^V=n_T$, estimating the amplitude $\bar{A}_{\rm CGWB}^V$, marginalizing over all the other parameters. Since the background is isotropic, the marginalized covariance of the amplitude of the circular polarization is 
\begin{equation}
    \begin{split}
  \sigma^2_{\bar{A}^V} =& \frac{1}{T_{\rm obs}}\frac{2}{\sum_f {\rm Tr}\left(\tilde{\gamma}^V_{f}S_f^{-1}\tilde{\gamma}^V_{f}S_f^{-1}\right)}+\,  \\ &\frac{\sum_{\hat{n},\hat{l},f,f^\prime}{\rm Tr}\left(\tilde{\gamma}^V_{f}S_f^{-1}B^V_{f,\hat{n}}S_f^{-1}\right) {\rm Tr}\left(\tilde{\gamma}^V_{f^\prime}S_{f^\prime}^{-1}B^V_{f^\prime,\hat{l}}S_{f^\prime}^{-1}\right)}{\left[\sum_{f^{\prime\prime}} {\rm Tr}\left(\tilde{\gamma}^V_{f^{\prime\prime}}S_{f^{\prime\prime}}^{-1}\tilde{\gamma}^V_{f^{\prime\prime}}S_{f^{\prime\prime}}^{-1}\right)\right]^2}\\
  &\times\frac{C^{VV}_{ff^\prime,\hat{n}\hat{l}}}{f^{n_T}f^{\prime\, \, n_T}}\, ,
  \label{eq:cov_cgwb_est}
  \end{split}
\end{equation}
where $\tilde{\gamma}^{V,AB}_f \equiv \sum_{\hat{n}}B_{f,\hat{n}}^{V,AB}$ is proportional to the overlap reduction function of the interferometer pair $(A,B)$. The tensor $B$ is defined as in Eq. \eqref{eq:def_B_V}, but with $f^{n_T}$ instead of $f^{2/3}$. The first term is the standard one computed, for instance, in~\cite{Allen:1996gp,Cornish:2001hg,Mentasti:2020yyd,Alonso:2020rar,LISACosmologyWorkingGroup:2022kbp}, while the second one is the covariance due to the component separation evaluated for the first time for the anisotropic case in~\cite{ValbusaDallArmi:2022htu}. It is clear that when $n_T\rightarrow 2/3$ and when $C_{ff^\prime,\hat{n}\hat{l}}^{VV}$ is independent of $\hat{n}$ and $\hat{l}$ (i.e., monopole signal), the CGWB and the AGWB are indistinguishable and the second contribution is proportional just to the amplitude of the AGWB fluctuations and it looks like the cosmic variance term computed in~\cite{Ricciardone:2021kel}, $\sigma^2_{\bar{A}^V}\sim \sigma^2_{\rm instr}+|A_{\rm foreground}^V|^2$. In our case, the AGWB is not isotropic, but flat in multipole space. This means that the factor that multiplies $C_{ff^\prime,\hat{n}\hat{l}}^{VV}$ is much smaller than one also in the case $n_T\rightarrow 2/3$. This is equivalent to say that measurements of the first few multipoles allow to break the degeneracy between the AGWB and the CGWB monopoles. The covariance due to the component separation is proportional to $1/T_{\rm obs}$, which implies that it scales in the same way of instrumental noise. We have found indeed that, although the detection of the circular polarization of the CGWB is mainly limited by instrumental noise, for any observing time the SNR receives corrections from the presence of the astrophysical foreground of $\sim 20\%$ when $n_T^V$ is approximately larger than minus one and smaller than three\footnote{When the filter applied to the data is chosen to maximize the amount of information from the shot noise, the shot noise is about twice the error due to instrumental noise, while in this case the filter used is different, thus the shot noise is smaller than the uncertainty associated to the detector response.}. More specifically, the square root of the covariance computed in Eq. \eqref{eq:cov_cgwb_est} is reduced by the $20\%$ when the astrophysical foreground is not considered, which means that neglecting the circular polarization of the AGWB computed in this work would lead to an underestimated error on the reconstruction of the circular polarization of the CGWB.

\paragraph{Conclusions}

Usually, measurements of the Stokes parameters of the SGWB are considered an efficient way to separate signals that overlap at interferometers, since the homogeneous and isotropic part of the astrophysical signal has a vanishing circular polarization, contrary to some cosmological models that can show up a polarized background. In this \textit{letter} we have shown that a non-negligible amount of circular polarization is present also in the AGWB, induced by the anisotropic distribution of the sources. In particular, we have focused on the AGWB generated by BBH mergers, for which the Poisson fluctuation of the number of sources, characterized by a flat angular power spectrum in the multipole space, generates a net amount of circular polarization, with a $f^{2/3}$ trend, which is sufficiently large that it is present at the level $\rm SNR \simeq 2.2$ at the network ET+CE for any observing time and must be
included in the noise model for these analyses. Although the significance of the detection does not improve with $T_{\rm obs}$, any increase in the number of detectors considered in the analysis would increase the SNR, because the outputs of the detectors in the network are correlated. In addition, the discussion we have done in this work is valid for any distribution of sources, thus it could be easily extended to stochastic backgrounds emitted in different frequency bands or for terrestrial detectors with different sensitivities. Furthermore, the circular polarization of the AGWB constitutes a troublesome foreground that limits the detection of a stochastic background of cosmological origin. In this preliminary work we have evaluated the efficiency of the subtraction of the circular polarization of the AGWB for ET+CE in the presence of a cosmological signal described by a power law. By exploiting the different features of the cosmological and astrophysical spectra, such as the frequency dependence and the features in the sky maps, it is possible to reconstruct the primordial background with very good precision, being limited just by instrumental noise as for the case of the intensity. The error associated to the component separation of the cosmological and the astrophysical background is not very large because of the significant difference between the sky maps of the CGWB we have considered (isotropic) and the AGWB map (white noise in the angular space), however it gives corrections of the order of $20\%$ to the total SNR, thus it has to be carefully taken into account. 

\paragraph{Acknowledgments.}
\noindent
L.V.  thanks the U. of Tokyo for the warm hospitality during the development of this work and the support from JSPS KAKENHI Grants No. JP20H04726. A.~N. is supported by JSPS KAKENHI Grants No. JP19H01894 and No. JP20H04726 and by Research Grants from the Inamori Foundation. A.~R. acknowledges financial support from the Supporting TAlent in ReSearch@University of Padova (STARS@UNIPD) for the project “Constraining Cosmology and Astrophysics with Gravitational Waves, Cosmic Microwave Background and Large-Scale Structure cross-correlations’'. S.M. acknowledges partial financial support by ASI Grant No. 2016-24-H.0.

\vskip 1.5cm

\onecolumngrid

\section{Population of Binary Black Holes}
BBHs are characterized by their intrinsic properties $\vec{\theta}$, by the redshift of the merger $z$ and by the time delay between the formation of the binary and the merger of the two objects\footnote{For practical purposes, here we include in the description of the population the detector window function $w$, thus the population discussed here is not the full BBH population, but the one that contributes to the astrophysical background.}. In our work we assume that we can factor out the dependence of the binary parameters w.r.t. the redshift and the time delay, 
\begin{equation}
\frac{\ud \bar{N}_{\rm GW}}{\ud t_d\ud\vec{\theta}\ud z}( t_d,\vec{\theta},z) = p(\vec{\theta})\frac{\ud R(t_d,z)}{\ud t_d}T_{\rm obs}\frac{\ud V}{\ud z\ud \Omega_e}(z_d)w(\vec{\theta},z)\, ,
\label{eq:def_N_GW_tot}
\end{equation}
where $T_{\rm obs}$ is the observation time, while the comoving volume element is defined by
\begin{equation}
\frac{\ud V}{\ud z\ud \Omega_e}(z_d) = \frac{c\, \chi^2(z_d)}{H(z_d)}\, ,
\end{equation}
with $\chi$ comoving distance. Note that however this is not true in general and one should use a generic, non-factorable expression, derived from the posterior of the observation of resolved sources at interferometers. Furthermore, we assume that the distribution of the astrophysical parameters that characterize the binary can be factorized in the following way,
\begin{equation}
p(\vec{\theta}) = p(m_1,m_2)p(\chi_1,\chi_2)p(\iota )\, .
\end{equation}
We use the Power-Law + Peak mass distribution for BHs of masses between $m_{\rm min}=2.5 \, M_\odot$ and $m_{\rm max}=100\, M_\odot$, and a uniform distribution in the mass ratio $q\equiv m_2/m_1$, with $q\leq 1$~\cite{LIGOScientific:2020kqk,LIGOScientific:2021psn}. We assume that the spin distribution of the compact objects is a Gaussian centered in $0$ and covariance $0.1$, while for the inclination angle we are considering an isotropic distribution of the sources, therefore we assume that $\cos\iota $ is drawn from a uniform PDF, thus
\begin{equation}
p(\iota )=p(\cos\iota )\frac{d\cos\iota }{d(-\iota) }= \frac{1}{2}\sin\iota\, \quad {\rm for} \quad 0 \leq \iota \leq \pi.
\end{equation}
The merger rate has been derived from the Star Formation Rate (SFR) at a given redshift, by using 
\begin{equation}
R_\star(z) = \int dM_h \, \langle {\rm SFR}(M_h,z) \rangle \frac{d\bar{N}_h}{dM_h}(M_h,z) \hspace{0.5em} [\rm yr^{-1}\,Mpc^{-3}]\, ,
\label{eq:R_star}
\end{equation}
where we have used the average SFR per halo from Universe Machine~\cite{Behroozi:2019kql} as in~\cite{Bellomo:2021mer} and the halo mass function of~\cite{Tinker:2008ff}. We connect the merger rate to the SFR by using
\begin{equation}
R(z) = R_{\rm LVK}(z=0)\frac{\int dt_d p(t_d) R_\star[z_d(z,t_d))]}{\int dt_d p(t_d) R_\star[z_d(z=0,t_d))]}\, ,
\end{equation}
where the LVK normalization considered has been taken from the central value of (for BBH), $R_{\rm LVK}(z=0) = 23.9^{+14}_{-8.6} \, \rm Gpc^{-3}\, yr^{-1}\, $~\cite{LIGOScientific:2021psn}. The time delay distribution has been taken as an inverse power law from $50 \, \rm Myr$ to the age of the Universe~\cite{Mapelli:2017hqk,Mapelli:2019bnp}. It is therefore straightforward to see that the merger rate for a given time delay is 
\begin{equation}
    \frac{\ud R(t_d,z)}{\ud t_d}\equiv  R_{\rm LVK}(z=0)\frac{p(t_d) R_\star[z_d(z,t_d))]}{\int dt_d p(t_d) R_\star[z_d(z=0,t_d))]}\, .
    \label{eq:dR_dtd}
\end{equation}
By combining Eqs. \eqref{eq:def_N_GW_tot}, \eqref{eq:dR_dtd}, it is straightforward to see that the mean number of GW events per halo is 
\begin{equation}
\begin{split}
    \bar{N}_{{\rm GW}|h}(M_{h},t_{d},\vec{\theta},z) = & \frac{R_{\rm LVK}(0)\langle {\rm SFR}(M_{h},t_{d},z)\rangle}{\int dt_d p(t_d) R_\star[z_d(z=0,t_d))]}\\
 &p(t_{d})p(\vec{\theta})w(\vec{\theta},z)T_{\rm obs}\frac{\ud V}{\ud z\ud \Omega_e}(z) \, \\
 &\Delta t_{d} \, \Delta \vec{\theta} \, \Delta z \, .
    \label{eq:def_bar_N_theta_td}
\end{split}
\end{equation}

\section{Shot-Noise Computation}
We are looking at the fluctuations of the object $\ud N_{\rm GW}/\ud t_d \ud \vec{\theta}$, which is the number of GW event at $z$ with parameters $\vec{\theta}$ and time delay $t_d$. We describe this events by using
\begin{equation}
    \frac{\ud N_{\rm GW}}{\ud t_d \ud \vec{\theta}\ud z}(\hat{n},t_{d},\vec{\theta},z) = \sum_{M_h}\frac{\ud N_{\rm GW}}{\ud M_h \ud t_d \ud \vec{\theta}\ud z}(\hat{n},M_{h},t_{d},\vec{\theta},z)\Delta M_{h}\, , 
\end{equation}
where we have defined the discrete summation in the following way,
\begin{equation}
    \sum_{M_h}\frac{\ud N_{\rm GW}(M_{h})}{\ud M_h \ud t_d \ud \vec{\theta}\ud z}\Delta M_{h} \equiv  \sum_{i=1}^{i_{\rm max}} \frac{\ud N_{\rm GW|h}(M_{h}^{(i)})}{\ud M_h \ud t_d \ud \vec{\theta}\ud z}\Delta M_{h}^{(i)} \, ,
    \label{eq:def_discrete_index_summation}
\end{equation}
and we have defined the number of GW events per each bin in all the halos of mass $M_h$ as
\begin{equation}
    \frac{\ud N_{\rm GW}(M_{h})}{\ud M_h \ud t_d \ud \vec{\theta}\ud z}\Delta M_{h}\equiv  \sum_{i=1}^{N_h(\hat{n},M_{h},t_{d},z)} \frac{\ud N^i_{\rm GW|h}}{\ud t_d \ud \vec{\theta}\ud z}(M_{h})\, .
\end{equation}
Both the number of GW events per halo and the number of halos with a given mass fluctuate according to a Poisson distribution,
\begin{equation}
\begin{split}
p(N^i_{{\rm GW}|h},\bar{N}_{{\rm GW}|h}) =& \frac{\bar{N}_{{\rm GW}|h}^{N^i_{{\rm GW}|h}}e^{-\bar{N}_{{\rm GW}|h}}}{N^i_{{\rm GW}|h}!}\, , \\
p(N_h,\bar{N}_h) =& \frac{\bar{N}_h^{N_h}e^{-\bar{N}_h}}{N_h!}\, ,
\end{split}
\end{equation}
where the mean number of GW events per halo has been computed in Eq. \eqref{eq:def_bar_N_theta_td}, while the average number of halos is simply defined by 
\begin{equation}
    \bar{N}_h(M_{h},t_{d},z) = \frac{\ud \bar{N}_h}{\ud M_h}(M_{h},t_{d},z) \Delta M_{h}\, ,
\end{equation}
with $\ud \bar{N}_h/\ud M_h$ the halo mass function. Note that $p(\vec{\theta})$ contains also information about the orientation angle of the binary. $N_{\rm GW}$ is obtained by the sum of many Poisson variables, therefore it follows a Compound Poisson Distribution~\cite{Jenkins:2019nks,Jenkins:2019uzp,Alonso:2020mva}. The expectation value of the CPD is the product of the expectation values of the two Poisson distributions,
\begin{equation}
\begin{split}
\bar{N}_{\rm GW}(t_{d},\vec{\theta},z)=&\sum_{M_h} \bar{N}_{h}(M_{h},t_{d},z),\bar{N}_{\rm GW|h}(M_{h},t_{d}, \vec{\theta}, z)\, .
\end{split}
\end{equation}
This expectation value has been used to compute the homogeneous and isotropic contribution to the background, which gives zero for the circular polarization . In this section we are interested in evaluating the impact of the fluctuation of $N_{\rm GW}$ to the signal. It is possible to show~\cite{Jenkins:2019nks,Jenkins:2019uzp,Alonso:2020mva} that the fluctuation of this shot-noise contribution is equal to 
\begin{equation}
\begin{split}
\sigma^2_{N_{\rm GW} N_{\rm GW}^\prime}\equiv & {\rm cov}\left[N_{\rm GW}(\hat{n},t_{d},\vec{\theta},z),N_{\rm GW}(\hat{n}^\prime,t_{d}^\prime,\vec{\theta}^\prime,z^\prime)\right] = \sum_{M_h}\bar{N}_{h}\left(\bar{N}_{\rm GW|h}+\bar{N}^2_{\rm GW|h}\right)  \delta_{\hat{n} \hat{n}^\prime}\delta_{\rm t_d t_d^\prime}\delta_{\vec{\theta} \vec{\theta}^\prime}\delta_{z z^\prime} \, ,
\end{split}
\end{equation}
where we have introduced the short notation for the Kronecker delta,
\begin{equation}
\begin{split}
    \begin{cases}
    X \equiv x_i \\
    X^\prime \equiv x_j
    \end{cases} \rightarrow
    \delta_{X X^\prime} \equiv \delta_{ij}\, . 
\end{split}
\end{equation}
On average the shot-noise fluctuation does not contribute to the AGWB signal, but since the covariance is not zero a net AGWB Stokes parameter is produced by a random realization
of the Compound Poisson Distribution,
%\begin{widetext}
\begin{equation}
    \begin{split}
        C^{\alpha \alpha^\prime}_{f f^\prime,\hat{n}\hat{n}^\prime}\equiv&{\rm cov}\left[\Omega_{\rm AGWB}^\alpha(\hat{n},f),\Omega_{\rm AGWB}^{\alpha^\prime}(\hat{n}^\prime,f^\prime)\right] = \\  =&\sum_{t_d,\vec{\theta},z}\sum_{t_d^\prime,\vec{\theta}^\prime,z^\prime}\frac{1}{(1+z)H(z)}\frac{1}{(1+z^\prime)H(z^\prime)}\frac{ ff^\prime}{(\rho_{\rm crit}c^2)^2}\frac{\ud E^V}{\ud f_e \ud\Omega_e}\left(\vec{\theta},z,f\right)\frac{\ud E^V}{\ud f_e \ud\Omega_e}\left(\vec{\theta},z,f^\prime\right)\frac{\sigma^2_{N_{\rm GW}N_{\rm GW}^\prime}}{T^2_{\rm obs}\frac{\ud V}{\ud z\ud \Omega_e}(z)\frac{\ud V}{\ud z\ud \Omega_e}(z^\prime)}= \\
        =&\sum_{t_d,\vec{\theta},z}\frac{1}{\left[(1+z)H(z)\right]^2}\frac{\frac{\ud E^\alpha}{\ud f_e \ud\Omega_e}\left(\vec{\theta},z,f\right)\frac{\ud E^{\alpha^\prime}}{\ud f_e \ud\Omega_e}\left(\vec{\theta},z,f^\prime\right)}{\left[T_{\rm obs}\frac{\ud V}{\ud z\ud \Omega_e}(z)\right]^2}\frac{ff^\prime}{(\rho_{\rm crit}c^2)^2}\\
        &\hspace{3em}\bar{N}_{h}(M_{h}, t_{d}, z)\left[\bar{N}_{\rm GW|h}(M_{h},t_{d},\vec{\theta},z)+ \bar{N}^2_{\rm GW|h}(M_{h},t_{d},\vec{\theta},z)\right]\delta_{\hat{n}\hat{n}^\prime} \, .
    \end{split}
\end{equation}
%\end{widetext}
We note that the $\bar{N}_{\rm GW|h}^2$ term is always subdominant, independently of the width of the bins, because of the low number of GW events per single halo. From now on we will therefore focus our analysis just on the linear term, which is analogous to the covariance obtained by a simple Poisson distribution.

\section{Computation of the covariance induced by foreground subtraction}

In this section we discuss a strategy that could be used to detect the circular polarization of a cosmological background in presence of the astrophysical foreground we have characterized in this work. The quantity that can be observed at the interferometer $A$ in a time segment $T$ is 
\begin{equation}
    d_{A,T}(t,f) \simeq \sum_{\hat{n}} \Delta\Omega\,  \sum_\alpha F_{A,f\hat{n}}^\alpha\, \left(h_{\alpha,f\hat{n}}^{\rm AGWB}+h_{\alpha,f\hat{n}}^{\rm CGWB}\right)+n_{A,T}(f)\, ,
\end{equation}
with $\Delta\Omega$ the solid angle of the pixels in which we have divided the sky, $F_A^\alpha$ the detector pattern function for the polarization $\alpha$ and $n_{A,T}$ the realization of the noise of the interferometer. The covariance of the data is therefore given by 
\begin{equation}
    \langle d_{f,A} \, d^*_{f^\prime,B}\rangle = \frac{1}{2}\frac{\delta_{f f^\prime}}{\Delta f}\left[N^{AB}_f+\sum_{\hat{n}}\sum_\alpha\tilde{B}^{\alpha,AB}_{f \hat{n}}\left(\Omega_{{\rm AGWB},\hat{n}f}^{\alpha}+\Omega_{{\rm CGWB},\hat{n}f}^{\alpha}\right)\right]\, ,
\end{equation}
where $N_f^{AB}$ is the power spectral density of the noise for the detector pair $(A,B)$ and we have defined
\begin{equation}
    \begin{split}
    \tilde{B}_{f,\hat{n}}^{{\rm I},AB} \equiv& \Delta \Omega \frac{3H_0^2}{4\pi^2 f^3}\left( F_{A,f\hat{n}}^+ F_{B,f\hat{n}}^{+\,*}+F_{A,f\hat{n}}^\times F_{B,f\hat{n}}^{\times\,*}\right)\, , \\
    \tilde{B}_{f,\hat{n}}^{{\rm V},AB} \equiv& \Delta \Omega \frac{3H_0^2}{4\pi^2 f^3}  \left( F_{A,f\hat{n}}^+ F_{B,f\hat{n}}^{\times\,*}-F_{A,f\hat{n}}^\times F_{B,f\hat{n}}^{+\,*}\right)\,i \, . 
    \end{split}
\end{equation}
Here we consider the simplified situation in which the cosmological background is homogeneous and isotropic, the tensor tilt of the cosmological background is known and we neglect the leakage of the intensity map into the reconstructed map of the circular polarization. We quantify the amount of circular polarization of the CGWB by using
\begin{equation}
    \Omega^V_{{\rm CGWB},\hat{n}f} = \bar{A}_{\rm CGWB}^V\left(\frac{f}{1\, \rm Hz}\right)^{n_T^V}\, ,
\end{equation}
where $\bar{A}_{\rm CGWB}^V$ is the unknown amplitude we are interested in. Since the amplitude is proportional to the square of the strain of the CGWB, we use an estimator quadratic in the data, 
\begin{equation}
    \hat{A}^V_{{\rm CGWB}} = \sum_{A,B,f,f^\prime}d_{f,A}E_{AB}^{f f^\prime}d_{f,B}^* - b \, ,
\end{equation}
where the bias $b$ can be found by imposing that the estimator is unbiased, $\langle \hat{A}^V_{{\rm CGWB}}\rangle = \bar{A}_{\rm CGWB}$, while the weights $E_{AB}^{ff^\prime}$ are chosen in order to minimize the covariance associated to $\hat{A}^V_{{\rm CGWB}}$. In~\cite{ValbusaDallArmi:2022htu} it has been shown that this covariance is the sum of two terms: the first contribution is proportional to fluctuations in the observed data, $\langle d^4\rangle -\langle d^2 \rangle^2$, while the second term is the error associated to the component separation between the CGWB and the AGWB. The computation of the estimator with the minimum covariance has been done in~\cite{ValbusaDallArmi:2022htu} and the covariance associated to the estimate of the amplitude of the CGWB is given by
\begin{equation}
\begin{split}
  \sigma^2_{\bar{A}^V} =& \frac{1}{T_{\rm obs}}\frac{2}{\sum_f {\rm Tr}\left(\tilde{\gamma}^V_{f}S_f^{-1}\tilde{\gamma}^V_{f}S_f^{-1}\right)}+\frac{\sum_{\hat{n},\hat{l},f,f^\prime}{\rm Tr}\left(\tilde{\gamma}^V_{f}S_f^{-1}B^V_{f,\hat{n}}S_f^{-1}\right) {\rm Tr}\left(\tilde{\gamma}^V_{f^\prime}S_{f^\prime}^{-1}B^V_{f^\prime,\hat{l}}S_{f^\prime}^{-1}\right)}{\left[\sum_{f^{\prime\prime}} {\rm Tr}\left(\tilde{\gamma}^V_{f^{\prime\prime}}S_{f^{\prime\prime}}^{-1}\tilde{\gamma}^V_{f^{\prime\prime}}S_{f^{\prime\prime}}^{-1}\right)\right]^2}\frac{C^{VV}_{ff^\prime,\hat{n}\hat{l}}}{f^{n_T}f^{\prime\, \, n_T}}\, ,
  \end{split}
\end{equation}
where we have defined
\begin{equation}
    B^V_{f,\hat{n}}\equiv \tilde{B}^V_{f,\hat{n}} \left(\frac{f}{1\, \rm Hz}\right)^{n_T^V}\, .
\end{equation}

\twocolumngrid

\end{document}